\documentclass[twocolumn,showpacs,preprintnumbers,amsmath,amssymb]{revtex4}

\usepackage{graphicx}
\usepackage{dcolumn}
\usepackage{bm}

\begin{document}

\preprint{APS}

\title{Reexamination of reaction rates for a key stellar reaction of $^{14}$O($\alpha$,p)$^{17}$F}

\author{J.J. He$^1$}
\email{jianjunhe@impcas.ac.cn}
\author{H.W. Wang$^2$}
\author{J. Hu$^{1,3}$}
\author{L. Li$^{1,3}$}
\author{L.Y. Zhang$^{1,3}$}
\author{M.L. Liu$^{1}$}
\author{S.W. Xu$^{1}$}
\author{X.Q. Yu$^{1}$}
\affiliation{$^1$Institute of Modern Physics, Chinese Academy of Sciences, Nanchang Road 509, Lanzhou 730000, China}%
\affiliation{$^2$Shanghai Institute of Applied Physics, Chinese Academy of Sciences, Shanghai 201800, China}
\affiliation{$^3$Graduate School of Chinese Academy of Sciences, Beijing 100049, China}

\date{\today}

\begin{abstract}
The reaction rates of the key stellar reaction of $^{14}$O($\alpha$,p)$^{17}$F have been reexamined. The previous conclusion, the 6.15-MeV state ($J^{\pi}$=1$^-$)
dominating this reaction rate, has been overthrown by a careful reanalysis of the previous experimental data [J. G\'{o}mez del Campo {\it et al.}, Phys. Rev. Lett.
{\bf 86}, 43 (2001)]. According to the present $R$-matrix analysis, the previous 1$^-$ assignment for the 6.15-MeV state is definitely wrong. Most probably, the
6.286-MeV state is the 1$^-$ state and the 6.15-MeV state is a 3$^-$ one, and hence the resonance at $E_x$=6.286 MeV ($J^{\pi}$=1$^-$) dominates the reaction rates in
the temperature region of astrophysical interests. The newly calculated reaction rates for the $^{14}$O($\alpha$,p)$^{17}$F reaction are quite different from the
previous ones, for instance, it's only about 1/6 of the previous value around 0.4 GK, while it's about 2.4 times larger than the previous value around 2 GK. The
astrophysical implications have been discussed based on the present conclusions.
\end{abstract}


\maketitle

Explosive hydrogen and helium burning are thought to be the main source of energy generation and a source for the nucleosynthesis of heavier elements in cataclysmic
binary systems, for example, x-ray bursters, {\em etc.}~\cite{bib:woo76,bib:cha92,bib:wie98}. During an x-ray burst (a high temperature and high density astrophysical
site), Hydrogen and Helium rich material from a companion star form an accretion disk around the surface of a neutron star where the H and He transferred from the disk
begin to pile up. The $\alpha$$p$ chain is initiated through the reaction sequence
$^{14}$O($\alpha$,$p$)$^{17}$F($p$,$\gamma$)$^{18}$Ne($\alpha$,$p$)$^{21}$Na~\cite{bib:bar00}, and increase the rate of energy generation by 2 orders of
magnitude~\cite{bib:wie98}. In x-ray burster scenarios, the nucleus $^{14}$O($t_{1/2}$=71 s) forms an important waiting point, and the ignition of the
$^{14}$O($\alpha$,$p$)$^{17}$F reaction at temperatures $\sim$0.4 GK produces a rapid increase in power and can lead to breakout from the hot CNO cycles into the
$rp$-process with the production of medium mass proton-rich nuclei~\cite{bib:sch98,bib:sch01,bib:bre09}. Excepting the $^{15}$O($\alpha$,$\gamma$)$^{19}$Ne reaction,
this reaction is arguably the most important reaction to be determined for x-ray burster scenarios.

Wiescher {\em et al.}~\cite{bib:wie87} calculated the reaction rates of the $^{14}$O($\alpha$,p)$^{17}$F reaction, and shown that the resonant reaction rates dominated
the total rates above temperature ~0.4 GK. However, Funck {\em et al.}~\cite{bib:fun88,bib:fun89} found that direct-reaction contributions to the $\ell$=1 partial wave
are comparable to or even greater than the resonant contributions at certain temperatures. Because the resonant reaction rates of $^{14}$O($\alpha$,p)$^{17}$F depend
sensitively on the excitation energies, spins, partial and total widths of the relevant resonances in $^{18}$Ne, Hahn {\em et al.}~\cite{bib:hah96} extensively
studied the levels in the compound system $^{18}$Ne~\cite{bib:hah96} by several reactions, such as, $^{16}$O($^3$He,$n$)$^{18}$Ne,$^{12}$C($^{12}$C,$^6$He)$^{18}$Ne
as well as $^{20}$Ne($p$,$t$)$^{18}$Ne reactions. Based on the firmer experimental results, they concluded that this reaction rate, in the important temperature regime
$\sim$0.5-1 GK, was dominated by reactions on a single 1$^-$ resonance at an excitation energy of 6.150 MeV lying 1.036 MeV above the $^{14}$O+$\alpha$ threshold
energy of 5.114 MeV. Harss {\em et al.}~\cite{bib:har99} studied the time reverse reaction $^{17}$F($p$,$\alpha$)$^{14}$O in inverse kinematics with $^{17}$F beam at
Argonne, and identified three levels at 7.16, 7.37, 7.60 MeV and determined their resonance strengths as well. Later, G\'{o}mez del Campo {\em et al.}~\cite{bib:gom01}
used the $p$($^{17}$F,$p$) resonant elastic scattering on a thick CH$_2$ target to look for resonances of astrophysical interest in $^{18}$Ne at ORNL. In the region
investigated, they located four resonances at excitation energies of 4.52, 5.10, 6.15, and 6.35 MeV in $^{18}$Ne, and J$^{\pi}$=1$^-$, 2$^-$ were respectively assigned
to the last two states based on their $R$-matrix analysis. Subsequently, Harss {\em et al.}~\cite{bib:har02} extracted the resonance strength and the width
$\Gamma_\alpha$ for the 6.15-MeV state based on this 1$^-$ assignment together with the excitation function obtained from their previous work~\cite{bib:har99}.
Recently, the inelastic component of this key 1$^-$ resonance in the $^{14}$O($\alpha$,p)$^{17}$F reaction has been studied by a new highly sensitive technique at
ISOLDE/CERN~\cite{bib:hjj09}, and found that this inelastic component will enhance the reaction rate, contributing approximating equally to the ground-state component
of the reaction rate, however not to the relative degree suggested in Ref.~\cite{bib:bla03}.

As a summary, all the previous discussions and calculations~\cite{bib:hah96,bib:har02,bib:bla03,bib:hjj09} related to the reaction rates of $^{14}$O($\alpha$,p)$^{17}$F
are based on the 1$^-$ assignment for the 6.15-MeV state. In this Letter, we completely overthrown this assignment by a carefully reanalyzing the experimental data
measured at ORNL~\cite{bib:gom01}, and the astrophysical consequences have been discussed based on the present new assignments.

\begin{figure}[b]
\includegraphics[width=8cm]{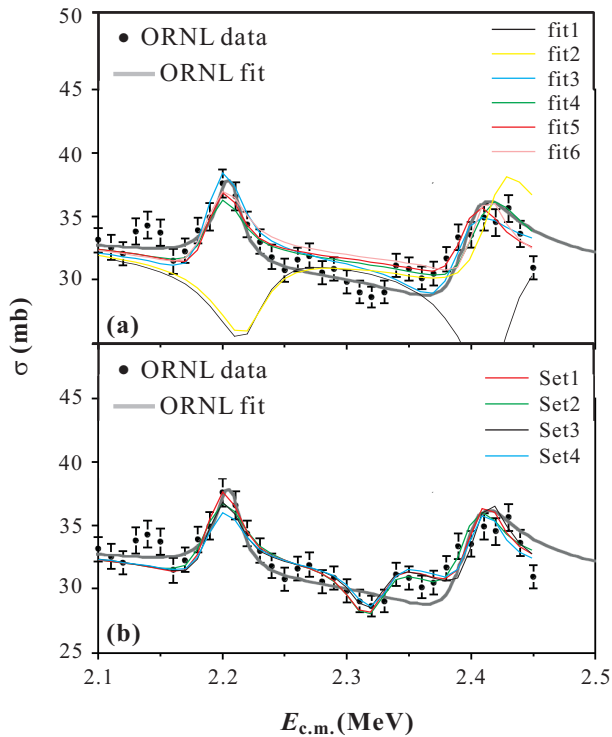}
\caption{\label{fig1} $R$-matrix fits for the experimental data measured at ORNL~\cite{bib:gom01}. The vertical scale corresponds to the angle integrated cross
sections in the range $\theta_\mathrm{CM}$=162$^\circ$$\sim$178$^\circ$. (a) fitting for two resonances, (b) fitting for three resonances. All curves are
convoluted by an assumed 10-keV energy resolution except those two labeled by `fit1' and `fit2'. For comparison, the ORNL fit is shown as well. See text for details.}
\end{figure}

\begin{table}
\caption{\label{table1}Resonant parameters used in FIG.\ref{fig1}. Here resonance energies ($E_r$) are in units of MeV, and proton partial widths ($\Gamma_p$) in keV.
The parameters in `fit1' and `fit2' were used in the previous work~\cite{bib:gom01}. See text for details.}
\begin{ruledtabular}
\begin{tabular}{|c|lll|lll|lll|}
 \multicolumn{1}{|c|}{}&\multicolumn{3}{c|}{Resonance 1} &\multicolumn{3}{c|}{Resonance 2} &\multicolumn{3}{c|}{Resonance 3}\\
 \cline{2-10}
Sets &  & & &  &  & &  &  & \\
     & $E_{r1}$ & $J^{\pi}$[$\ell,s$] & $\Gamma_p$ & $E_{r2}$ & $J^{\pi}$[$\ell,s$] & $\Gamma_p$ & $E_{r3}$ & $J^{\pi}$[$\ell,s$] & $\Gamma_p$\\
\hline
fit1\footnotemark[1] & 2.22 & 1$^-$[1, 2] & 50 & 2.42 & 2$^-$[1, 3] & 50 &      &             &     \\
fit2\footnotemark[1] & 2.22 & 1$^-$[1, 2] & 50 & 2.42 & 2$^-$[1, 2] & 50 &      &             &     \\
fit3\footnotemark[2] & 2.20 & 3$^-$[1, 3] & 15 & 2.39 & 3$^-$[1, 2] & 30 &      &             &     \\
fit4\footnotemark[2] & 2.20 & 2$^-$[1, 2] & 15 & 2.41 & 2$^-$[1, 2] & 30 &      &             &     \\
fit5\footnotemark[2] & 2.20 & 3$^-$[1, 3] & 10 & 2.40 & 2$^-$[1, 2] & 20 &      &             &     \\
fit6\footnotemark[2] & 2.20 & 2$^-$[1, 2] & 20 & 2.41 & 3$^-$[1, 3] & 10 &      &             &     \\
Set1\footnotemark[2] & 2.20 & 3$^-$[1, 3] & 12 & 2.40 & 2$^-$[1, 2] & 20 & 2.32 & 1$^-$[1, 2] & 15  \\
Set2\footnotemark[2] & 2.20 & 2$^-$[1, 2] & 20 & 2.40 & 2$^-$[1, 2] & 20 & 2.32 & 1$^-$[1, 2] & 15  \\
Set3\footnotemark[2] & 2.20 & 3$^-$[1, 3] & 10 & 2.41 & 3$^-$[1, 3] & 12 & 2.32 & 1$^-$[1, 2] & 15  \\
Set4\footnotemark[2] & 2.20 & 2$^-$[1, 2] & 15 & 2.41 & 3$^-$[1, 3] & 10 & 2.33 & 1$^-$[1, 2] & 12  \\
\end{tabular}
\footnotemark[1] no energy-resolution convolution in the fits of FIG.\ref{fig1};
\footnotemark[2] a 10-keV energy-resolution convolution in the fits of FIG.\ref{fig1}.
\end{ruledtabular}
\end{table}

\begin{figure}
\includegraphics[width=8cm]{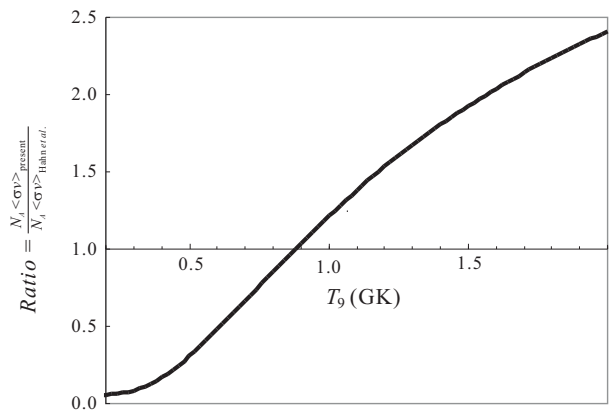}
\caption{\label{fig2} The ratios between the present resonant reaction rates and those previous ones at certain temperature range. See text for details.}
\end{figure}

In the present analysis, the multichannel $R$-matrix calculations~\cite{bib:lan58,bib:des03,bib:bru02}(see example~\cite{bib:alex}) that include the energies, widths,
spins, angular momenta, and interference sign for each candidate resonance have been performed with a channel radius of $r_0$=1.25 fm ($R$=$r_0$$\times$(1+17$^{1/3}$)),
appropriate for the $^{17}$F+$p$ system. The difference between the code we used and the {\tt MULTI} code~\cite{bib:nel85} utilized in the previous
work~\cite{bib:gom01} is negligibly small, which ensures the correctness of the present $R$-matrix analysis. The ground state spin-parity configurations of $^{17}$F
and the proton are 5/2$^+$ and 1/2$^+$, respectively, and therefore the channel spin in the elastic channel can have two values $s$=2, 3.

The fitting curve shown in the upper panel of Fig. 2 in~\cite{bib:gom01} has been exactly reproduced by utilizing their resonant parameters, {\em i.e.},
$E_{cm}$=0.6 MeV, 3$^+$, $\Gamma$=18 keV; $E_{cm}$=1.18 MeV, 2$^+$, $\Gamma$=45 keV (1.18 MeV was mistyped by 1.118 MeV in~\cite{bib:gom01}); and the last `dip'
structure can be reproduced by the following parameters: $E_{cm}$=1.53 MeV, 2$^-$, $\Gamma$=5 keV, which is consistent with the results of Ref.~\cite{bib:hah96}
($\Gamma$$\leq$20 keV). However, according to our $R$-matrix analysis, 1$^-$ assignment for the 6.15-MeV state is absolutely impossible (see fitting curves labeled by
`fit1' and `fit2' in FIG. 1(a)), the shape of the 1$^-$ resonance ($\ell$=1) is of a `dip' structure instead of a `bump' one. In addition, all possible combinations of
different spin-parity assignments for these two states have been attempted and the most probable fitting curves are shown in FIG. 1(a) with parameters listed in
Table 1. In order to achieve a better fit, all curves are convoluted by an assumed 10-keV energy resolution except those two labeled by `fit1' and `fit2', of course
this will not affect the spin-parity assignments for the resonances. Furthermore, we have tried to fit the experimental data shown in the lower panel of Fig. 2
in~\cite{bib:gom01} with three resonances, and the most probable fitting curves are shown in FIG. 1(b) with parameters listed in Table 1. It's very obvious this kind
of three-resonance fits reproduce the experimental data better than those two-resonance ones, especially the `dip' structure at $E_{cm}$=2.32 MeV can be fitted very
well.

The present $R$-matrix analysis shows that two states at $E_{cm}$=2.20($E_x$=6.12), 2.40($E_x$=6.32) both possibly have $J^{\pi}$=2$^-$ or 3$^-$, while the state at
$E_{cm}$=2.32($E_x$=6.24) most probably has $J^{\pi}$=1$^-$. These three states should correspond to the $E_x$=6.150, 6.345, and 6.286 MeV states observed
before~\cite{bib:hah96} within a $\sim$30 keV uncertainty. In order to constrain the spin-parity assignments for these states, let's examine the well-known mirror
nucleus $^{18}$O, which has only three known levels $J^{\pi}$=1$^-$,(2$^-$), and 3$^-$ around this energy region~\cite{bib:hah96}. The strong population of the
6.15-MeV state in the $^{16}$O($^3$He,$n$)$^{18}$Ne reaction~\cite{bib:hah96} suggests it has natural parity, which eliminates the possibility of $J^{\pi}$=2$^-$.
According to the above $R$-matrix analysis, it's absolutely not a 1$^-$ state, and therefore it should be a 3$^-$ state. The results from the
$^{12}$C($^{12}$C,$^6$He)$^{18}$Ne and $^{20}$Ne($p$,$t$)$^{18}$Ne reactions suggest that 6.286-MeV state is of natural parity and 6.345-MeV state of unnatural parity.
Therefore, we propose that these two states most probably have 1$^-$ and 2$^-$, respectively. Actually Funck {\em et al.}~\cite{bib:fun88} predicted a 1$^-$ state at
6.294 MeV. The ORNL experimental data can be reproduced very well with these assignments (see fitting curve labeled by `Set1' in FIG. 1(b)).

In the temperature region interesting for x-ray burster scenarios~\cite{bib:hah96,bib:har02,bib:hjj09}, only two natural-parity states, {\em i.e.},
$E_x$=6.150 ($J^{\pi}$=3$^-$, $\ell_\alpha$=3), 6.286 ($J^{\pi}$=1$^-$, $\ell_\alpha$=1), are needed to calculate the reaction rates of $^{14}$O($\alpha$,p)$^{17}$F.
Our new assignments for these two states are just contrary to the previous ones~\cite{bib:hah96}. The previously calculated $\Gamma_\alpha$ partial widths are 2.2, 0.34 eV for the 6.150, 6.286 MeV states, respectively. According to the relationship of
$\Gamma_\alpha \propto C^2S_\alpha \times P_\ell$($E_r$)~\cite{bib:hjj091}, the presently calculated $\Gamma_\alpha$ partial widths are 0.051, 13.4 eV
for these two states, respectively. Accordingly, the calculated resonant strength $\omega\gamma$$_{(\alpha,p)}$ are 0.36, 40 eV, respectively, while they were
6.6, 2.4 eV in the previous work~\cite{bib:hah96}. As a consequence, the roles of these two resonances in contributing the resonant reaction rate of
$^{14}$O($\alpha$,p)$^{17}$F are exchanged, and now the resonance at $E_x$=6.286 MeV ($J^{\pi}$=1$^-$,$\ell_\alpha$=1) dominates the total reaction rates within the
temperature region of interests (0.4$\sim$3 GK). The resonant reaction-rate ratios between the present results and the previous ones~\cite{bib:hah96} are plotted in
Fig.~\ref{fig2}, here only two resonances ($E_x$=6.150, 6.286 MeV) are included in the calculations. It can be seen that the present reaction rate is quite different
from the previous one, for instance, it's only about 1/6 of the previous value around 0.4 GK but, about 2.4 times larger than the previous one around 2 GK.
According to the present analysis, we think the 1$^-$ assignment for the 6.286-MeV state is quite reasonable, and hence the spin-parity assignments for the 6.150,
6.345 MeV states are rather unimportant in calculating the reaction rates ({\em i.e.}, whether they are $J^{\pi}$=2$^-$,3$^-$, or vice versa).

The present rates confirms that the $^{14}$O($\alpha$,p)$^{17}$F reaction is rather unlikely to be dominant component in the hot CNO cycles in novae environments
(instead $^{15}$O($\alpha$,$\gamma$)$^{19}$Ne reaction is, see discussions in~\cite{bib:hah96}). Due to the present rate enhancements above 0.9 GK, this reaction can,
however, contribute strongly to the breakout from the hot CNO cycle under the more extreme conditions in x-ray bursters. The present conclusion could probably affect
the onset temperature where the $\alpha$-capture dominates $\beta$-decay and a breakout from the hot CNO cycle via $^{14}$O($\alpha$,p)$^{17}$F reaction begins to take
place~\cite{bib:wie87}. In addition, the present conclusion shows that the previous inelastic-scattering contributions~\cite{bib:bla03,bib:hjj09} to the total reaction
rates of $^{14}$O($\alpha$,p)$^{17}$F could be neglected.
The detailed reaction rate calculations for this key reaction and its astrophysical implications will be published later~\cite{bib:wan10}.

\begin{acknowledgments}
This work is financially supported by the the ``100 Persons Project" and the ``Project of Knowledge Innovation Program"  of Chinese Academy of Sciences (KJCX2-YW-N32),
the National Natural Science Foundation of China(10975163, 10505026), and the Major State Basic Research Development Program of China (2007CB815000).
\end{acknowledgments}


\begin{thebibliography}{99}
\bibitem{bib:woo76}
S.E. Woosley, R.E. Taam, Nature {\bf 263}, 101 (1976).
\bibitem{bib:cha92}
A.E. Champage and M. Wiescher, Annu. Rev. Nucl. Part. Sci. {\bf 42}, 39 (1992).
\bibitem{bib:wie98}
M. Wiescher, H. Schatz, and A.E. Champagne, Phil. Trans. R. Soc. A {\bf 356}, 2105 (1998).
\bibitem{bib:bar00}
D.W. Bardayan, J.C. Blackmon, C.R. Brune {\it et al.}, Phys. Rev. C {\bf 62}, 055804 (2000).
\bibitem{bib:sch98}
H. Schatz, A. Aprahamian, J. G$\ddot{o}$rres {\it et al.}, Phys. Rep. {\bf 294}, 167 (1998).
\bibitem{bib:sch01}
H. Schatz, A. Aprahamian, V. Barnard {\it et al.}, Phys. Rev. Lett. {\bf 86}, 3471 (2001).
\bibitem{bib:bre09}
M. Breitenfeldt, G. Audi, D. Beck {\it et al.}, Phys. Rev. C {\bf 80}, 035805 (2009).
\bibitem{bib:wie87}
M. Wiescher, V. Harms, J. G$\ddot{o}$rres {\it et al.}, Astrophys. J. {\bf 316}, 162 (1987).
\bibitem{bib:fun88}
C. Funck, and K. Langanke, Nucl. Phys. {\bf A480}, 188 (1988).
\bibitem{bib:fun89}
C. Funck, B. Grund, and K. Langanke, Z. Phys. {\bf A332}, 109 (1989).
\bibitem{bib:hah96}
K.I. Hahn, A. Garc\'{i}a, E.G. Adelberger {\it et al.}, Phys. Rev. C {\bf 54}, 1999 (1996).
\bibitem{bib:har99}
B. Harss, J.P. Greene, D. Henderson {\it et al.}, Phys. Rev. Lett. {\bf 82}, 3964 (1999).
\bibitem{bib:gom01}
J. G\'{o}mez del Campo, A. Galindo-Uribarri, J.R. Beene {\it et al.}, Phys. Rev. Lett. {\bf 86}, 43 (2001).
\bibitem{bib:har02}
B. Harss, C.L. Jiang, K.E. Rehm {\it et al.}, Phys. Rev. C {\bf 65}, 035803 (2002).
\bibitem{bib:hjj09}
J.J. He, P.J. Woods, T. Davinson {\it et al.}, Phys. Rev. C {\bf 80}, 042801(R) (2009).
\bibitem{bib:bla03}
J.C. Blackmon, D.W. Bardayan, W. Bradfield-Smith {\it et al.}, Nucl. Phys. {\bf A718}, 127(c) (2003).
\bibitem{bib:lan58}
A.M. Lane and R.G. Thomas, Rev. Mod. Phys. {\bf 30}, 257 (1958).
\bibitem{bib:des03}
P. Descouvemont, {\it Theoretical Models for Nuclear Astrophysics} (Nova Science Pubishers Inc., New York, 2003).
\bibitem{bib:bru02}
C.R. Brune, Phys. Rev. C {\bf 66}, 044611 (2002).
\bibitem{bib:alex}
A.St.J. Murphy, A.M. Laird, C. Angulo, Phys. Rev. C {\bf 79}, 058801 (2009).
\bibitem{bib:nel85}
R.O. Nelson, E.G. Bilpuch, and G.E. Mitchell, Nucl. Instrum. Methods Phys. Res. Sect. A {\bf 236}, 128 (1985).
\bibitem{bib:hjj091}
J.J. He, S. Kubono, T. Teranish {\it et al.}, Phys. Rev. C {\bf 80}, 015801 (2009).
\bibitem{bib:wan10}
J.J. He, H.W. Wang, J. Hu {\it et al.}, under preparation.


\end{thebibliography}

\end{document}